\documentclass[prc,twocolumn,superscriptaddress,superscriptfootnote]{revtex4}
\usepackage{amssymb}

\usepackage{amsmath,amsfonts,amssymb,epsfig,color}

\newcommand{\Spn}[1]{\ensuremath{\mathrm{Sp}( #1 )}}

\newcommand{\so}[1]{\ensuremath{\mathfrak{so}( #1 )}}

\newcommand{\spn}[1]{\ensuremath{\mathfrak{sp}( #1 )}}

\newcommand{\IAS}{isobaric analog $0^+$ state}
\newcommand{\IASs}{isobaric analog $0^+$ states}
\newcommand{\fpg}{\ensuremath{1f_{5/2}2p_{1/2}2p_{3/2}1g_{9/2}} }

\newcommand{\flevel}{\ensuremath{1f_{\frac{7}{2}}} }
\newcommand{\dlevel}{\ensuremath{1d_{\frac{3}{2}}} }

\begin{document}

\title{Isospin Symmetry Breaking in an Algebraic Pairing Sp(4) Model}
\author{K. D. Sviratcheva}
\affiliation{Department of Physics and Astronomy, Louisiana State University,
Baton Rouge, LA 70803, USA}
\author{A. I. Georgieva}
\affiliation{Department of Physics and Astronomy, Louisiana State University,
Baton Rouge, LA 70803, USA}
\affiliation{Institute of Nuclear Research and Nuclear Energy,
Bulgarian Academy of Sciences, Sofia 1784, Bulgaria}
\author{J. P. Draayer}
\affiliation{Department of Physics and Astronomy, Louisiana State University,
Baton Rouge, LA 70803, USA}
\date{\today}

\begin{abstract}
An exactly solvable \spn{4} algebraic approach extends beyond the
traditional isospin conserving nuclear interaction to bring forward
effects of isospin symmetry breaking and isospin mixing resulting from a two-body
nuclear interaction that includes proton-neutron ($pn$) and like-particle
isovector pairing correlations plus significant isoscalar $pn$
interactions. The model yields an estimate for the extent to which \IASs~
in light and  medium mass nuclei may mix with one another and
reveals possible,  but still extremely weak, non-analog $\beta
$-decay transitions.
\end{abstract}

\maketitle

\section{Introduction}

A fundamental feature of nuclear structure is the basic symmetry between
neutrons and protons, namely, the {\it charge independence} of the
nuclear force, that is evident  in the striking similarity in the energy
spectra of  nuclear isobars \cite{BohrMottelson}. This  implies that the
proton-proton ($pp$) interaction and the neutron-neutron ($nn$)
interaction are equal to the isospin
$T =1$  $pn$ interaction  and leads to `rotational' invariance in
isotopic space. However, the isospin invariance is violated by  the
electromagnetic interaction, mainly the Coulomb repulsion between
nucleons, which has become the focus of many phenomenological and
microscopic studies
\cite{Wigner57,GoswamiChen,Auerbach72,BertschW73,BohrMottelson,TownerHH77,
Lawson,BenensonK79,DobaczewskiH95,OrmandBrown,SagawaGS96,
CivitareseReboiroVogel,NavratilBO97,HardyT02,Lisetskiy,
BesC04,AbergHMR04,MichelNP04,PetroviciSRF05,AlvarezRGS05}.

The primary effect of the Coulomb force is to introduce
into the  theory a dependence on the third isospin projection, $T_0$, resulting
in energy splitting of  isobaric analog states (a
$T$-multiplet) without coupling different isospin  multiplets.
At the same time, the isospin-violating part of the Coulomb 
interaction  leads to
small isospin mixing in
nuclear ground states increasing with $Z$ and largest for
$N=Z$. The ground state isospin impurity is theoretically estimated
to be as small as a
percent for nuclei in the \flevel level \cite{BohrMottelson}, and up
to $4-5\%$ toward the
$fpg$ shell closure \cite{DobaczewskiH95}. Another source of mixing
probability is the
isospin  non-conserving part of the
nuclear Hamiltonian, which includes effects due to the proton-neutron
mass difference and small charge dependent components in the  strong
nucleonic interaction \cite{Auerbach72}.
Experimental results clearly reveal the existence of  isospin mixing
\cite{Hagberg,Savard95}. The increase in isospin mixing towards medium
mass nuclei has been detected in novel high-precision experiments
\cite{Cottle99,Garrett01,Piechaczek03,Farnea03,Ekman04}, which
continue to push the exploration of unstable nuclei  with the advent of
advanced  radioactive beam facilities.

The violation of the charge independence of the nuclear
interaction is well
established. The purely nuclear parts of the $pp$ force and the $T
=1$ $pn$ force differ from
each other, which appears to be associated with the electromagnetic
structure of the nucleons
\cite{Auerbach72}. An analysis of the $^1S$ scattering in the $pn$
system and the low-energy
$pp$ scattering  lead to the estimate that the nuclear interaction
between protons and
neutrons ($V_{pn}^{T =1}$) in $T =1$ states are more attractive than
the force between the
protons ($V_{pp}$) by $2\%$, $|V_{pn}^{T =1}-V_{pp}|/V_{pp} \sim 2\%$
\cite{Henley66}. In addition, the charge asymmetry between the $pp$
and $nn$ interactions was
found to be smaller; namely, less than $1\% $ \cite{Baumgartner66}.
More recent investigations confirm isospin violation in light nuclei
\cite{NavratilBO97,NavratilVB00,MachleidtM01,PieperPWC01,NavratilC04} starting
with modern charge dependent realistic interactions
\cite{StoksKTS94,WiringaSS95,MachleidtSS9601,EntemM03} and including valuable
input information from particle physics (see for example,
\cite{CoonBlunden8287,BarnesCL03}).  Furthermore, after the Coulomb energy is
taken into account the discrepancy in the isobaric-multiplet energies is bigger
for the seniority zero  levels as compared to higher-seniority  states
indicating the presence of a short range charge dependent interaction
\cite{Lawson}. Indeed, the $J=0$ pairing correlations have been recently shown
to have an overwhelming dominance in the isotensor energy difference within
isobaric multiplets \cite{ZukerLMP02}, which manifests
itself in the
charge dependent $T=2$ nature of the pairing interaction.

The findings mentioned above point out the need for a charge dependent
microscopic description of $J=0$ pairing correlations.
An algebraic \spn{4} approach is ideally suited for this purpose
\cite{SGD03,SGD04IAS} for it combines, on the one hand a
microscopic modeling
of the pairing interaction and its charge dependence, and on
the other hand,
a straightforward scheme for estimating the significance of the isospin mixing
due to pairing correlations without the need for carrying out large-dimensional
matrix diagonalizations. Strong isospin breaking in pair formation, if found,
implies a significant presence of isospin admixture among the seniority-zero
\IASs~including $0^+$ ground states. This in turn will affect the predictive
power of precise studies of superallowed  $0^+ \rightarrow 0^+$ Fermi $\beta
$-decay transitions.  This is because the latter provide reliable tests of
isospin mixing (see \cite{TownerH02} for a review), and as well furnish a
precise test of the unitary condition of the Cabbibo-Kobayashi-Maskawa matrix
\cite{CKM63and73} (for a review of this subject, see \cite{TownerH03}).

Our objective is to explore isospin mixing beyond that due to  the Coulomb
interaction, which is isolated with the help of an advanced Coulomb correction
formula \cite{RetamosaCaurier}. Specifically, we focus on the isospin
non-conserving part  of the {\it pure} nuclear interaction, which recently has
been  found to be at least as important as the Coulomb repulsion
\cite{ZukerLMP02}. The  outcome of this study shows the significance of the
pairing charge dependence and its  role in mixing isospin multiplets of
pairing-governed \IASs.

\section{Theoretical Framework: the Reasonable Approximation}

We employ a simple but powerful group-theoretical model, which
is based on the \spn{4} algebra (isomorphic to \so{5}
\cite{Hecht,Ginocchio,EngelLV96}). The \Spn{4} microscopic model is precisely
suitable for the qualitative study of isospin violation in \IASs~ because it
naturally extends the isospin invariant nuclear interaction to incorporate
isospin non-conserving forces, while it retains the \Spn{4} dynamical  symmetry
of the Hamiltonian (see \cite{VanIsacker} for a review on dynamical 
symmetries).

A comparison with experimental data demonstrates that the
\Spn{4} model provides a reasonable description of the  pairing-governed \IASs~
\footnote{The lowest among these states include ground states for even-even
nuclei and only some  ($N\approx Z$) odd-odd nuclei, as well as, for example,
low-lying $0^+$ states in odd-odd nuclei that have the same isospin 
as the ground
state of a semi-magic even-even isobaric neighbor with fully-paired protons (or
neutrons). } in light and  medium mass nuclei, where  protons and neutrons
occupy the same shell \cite{SGD03,SGD03stg,SGD04IAS}. The two-body  model
interaction includes proton-neutron and like-particle pairing plus symmetry
terms and contains a non-negligible  implicit portion of the
quadrupole-quadrupole interaction
\cite{cmpRealInt05}. Moreover, the
\Spn{4} model interaction itself, which relates to the whole energy
spectrum rather than to a single $J^\pi =0^+$ $T=1$ state, was found to be
quite strongly correlated ($0.85$) with the realistic CD-Bonn+3terms
interaction \cite{PopescuSVN05} in the $T=1$ channel and with an overall
correlation of $0.76$ with the realistic GXPF1 interaction 
\cite{HonmaOBM04} for
the \flevel orbit \cite{cmpRealInt05}. In short, the relatively simple
\Spn{4}  model seems to be a reasonable approximation that reproduces 
especially
that part of  the interaction that is responsible for shaping pairing-governed
\IASs.

The \Spn{4} model reflects the symplectic dynamical symmetry of
\IASs~ \cite{SGD04IAS} determined by the strong nuclear interaction.
The weaker Coulomb
interaction breaks this symmetry and significantly complicates the
nuclear  pairing problem.
This is why, in our investigation we adopt a sophisticated
phenomenological Coulomb  correction  to the
experimental energies such that a nuclear system can be regarded as
if there is no  Coulomb
interaction between its constituents. The {\it Coulomb corrected}
experimental energy,
$E_{\exp }$, for given valence protons $N_{+1}$ and neutrons $N_{-1}$
is adjusted to be
\begin{eqnarray}
E_{\exp }(N_{+1},N_{-1})&=&E^C_{\exp }(N_{+1},N_{-1})- E^C_{\exp }(0,0)
\nonumber
\\ &+&V_{Coul}(N_{+1},N_{-1}),
\label{EexpCoul}
\end{eqnarray}
where \footnote{To avoid confusion we mention that in
(\ref{EexpCoul}) the energies are
assumed positive for bound states; $V_{Coul}$ is also defined positive.}
$E^C_{\exp }$ is the total measured energy including the Coulomb
energy, $E^C_{\exp }(0,0)$ is
the binding energy of the core, and
     $V_{Coul}(N_{+1},N_{-1})$ is the Coulomb correction for a nucleus
with mass $A$ and $Z$
protons taken relative to the core $V_{Coul}(N_{+1},N_{-1})=
V_{Coul}(A,Z)-V_{Coul}(A_{core},Z_{core})$. The recursion formula for
the $V_{Coul}(A,Z)$
Coulomb energy is derived in \cite {RetamosaCaurier} with the use of
the Pape and Antony formula \cite{PapeAntony88}.
The Coulomb corrected energies (\ref{EexpCoul}) should reflect solely
the nuclear properties
of the many-nucleon systems.

Assuming charge independence of the nuclear force, the general
isoscalar Hamiltonian with Sp(4) dynamical symmetry, which consists
of one- and two-body terms  and conserves the number of particles, can be
expressed  through the Sp(4) group generators,
\begin{eqnarray}
H_{0} =&-G\sum _{i=-1}^{1}\hat{A}^{\dagger }_{i}
\hat{A}_{i}-\frac{E}{2\Omega} (\hat{T}
^2-\frac{3\hat{N}}{4 })
\nonumber \\
&-C\frac{\hat{N}(\hat{N}-1)}{2}-\epsilon  \hat{N},
\label{clH0}
\end{eqnarray}
where $\hat{T}^2=\Omega \{ \hat{T}_+,\hat{T}_-\}+\hat{T}_0^2$ and
$2\Omega $ is the shell dimension for a given nucleon type. The
generators $\hat{T}_{\pm}$
and $\hat{T}_{0}$ are the valence isospin operators,
$\hat{A}^{(\dagger ) }_{0,+1,-1}$ create (annihilate) respectively a
proton-neutron ($pn$)
pair, a  proton-proton ($pp$) pair or a neutron-neutron
$(nn)$ pair of total angular momentum $J^{\pi}=0^+$ and isospin
$T=1$, and $\hat{N} =
\hat{N}_{+1}+\hat{N}_{-1}$ is the total number of valence particles
with an eigenvalue $n$.
The $G,E$ and $C$ are interaction strength parameters  and
$\epsilon >0$ is the Fermi level energy (see Table I in
\cite{SGD04IAS} for estimates). The isospin
conserving Hamiltonian
(\ref{clH0}) includes an isovector ($T=1$) pairing interaction
($G\geq  0 $ for attraction)
and a diagonal isoscalar ($T=0$) force, which is related to a
symmetry term ($E$).

Charge dependent but charge symmetric nucleon-nucleon interaction
($V_{pp}=V_{nn} \ne V_{pn}$)  brings into the nuclear
Hamiltonian a  small isotensor component (with zero third isospin
projection so that the
Hamiltonian commutes with
$T _0$). This is achieved in the framework of the \Spn{4} model by
introducing the two additional terms,
\begin{equation}
H_{\text{IM}}=-F \hat{A}^{\dagger }_{0}\hat{A}_{0}, \qquad
H_{\text{split}}=-D(\hat{T} _{0}^2-\frac{\hat{N}}{4}),
\label{HINC}
\end{equation}
to the isospin invariant model Hamiltonian (\ref{clH0}) in a way that
the Hamiltonian
\begin{equation}
H=H_{0}+H_{\text{IM}}+H_{\text{split}}
\label{clH}
\end{equation}
possesses \Spn{4} dynamical symmetry. In other words,
charge dependence is introduced into the model pairing Hamiltonian 
(\ref{clH0}) by
allowing the strength of two of the underlying interactions to vary.
The interaction strength parameters
$F$ and
$D$ (\ref{HINC}) determined in an optimum fit over a significant 
number of nuclei
(total of 149)
\cite{SGD03} are given in Table
\ref{tab:fitStat} and yield non-zero values. These parameters yield
quantitative  results that are superior than the ones with $F=0$ and $D=0$; for
example, in  the case of the
\flevel level the variance between the model and experimental energies of the
lowest
\IASs~ increases by $85\%$ when the $D$ and $F$ interactions are turned off.
For the  present investigation the parameters in (\ref{clH0}) along 
with $F$ and
$D$ (\ref{HINC}) are not varied as their values were fixed to be physically
valid and to yield reasonable energy
\cite{SGD03,SGD04IAS} and fine structure \cite{SGD03stg} reproduction for light
and medium mass nuclei with valence protons and  neutrons  occupying the same
shell. For these nuclei in the mass range $32\le A\le 100$,  the
pairing-governed \IASs~are well described, but still approximately, by the
eigenvectors of the effective Hamiltonian (\ref{clH}) in a basis of 
fully-paired
$0^+$ states \cite{SGD04IAS}.
\begin{table}[th]
\caption{Interaction strength parameters related to the isospin
problem for
three regions of nuclei  specified by the valence model space.
$F$, $D$, and $E$ are in MeV.}
\center{
\begin{tabular}{lccc}
\hline
\hline
\begin{tabular}{lr}
   &  \\
   & \text{Model} \\
\text{Strength} & \text{space} \\
\text{parameters} &
\end{tabular}
& $(\dlevel )$ & $(\flevel )$  &
                                     \begin{tabular}{c}
                                      $(1f_{5/2}2p_{1/2}$ \\
                                      $2p_{3/2}1g_{9/2})$
                                      \end{tabular} \\
\hline
$F/{\Omega }$      & 0.007 &  0.072 &  0.056  \\
$D$                & 0.127 &  0.149 & -0.307  \\
$\left|D/\frac{E}{2\Omega }\right|$   &0.090 & 0.133 & 0.628  \\
\hline \hline
\end{tabular}
}
\label{tab:fitStat}
\end{table}

While the second interaction ($H_{\text{split}}$) in (\ref{HINC})
takes into account only the
splitting of the isobaric analog energies, the first correction
induces small isospin mixing
(IM). The isospin mixing interaction (\ref{HINC}) does not account
for the entire interaction
that mixes states of same angular momentum and parity but  different
isospin values. It only describes a possible $\Delta T=2$ mixing between
\IASs~due to a pure nuclear pairing interaction. While the extent
of such isospin admixing is expected to be smaller than the total mixing due to
isospin non-conserving terms
\cite{TownerHH77,OrmandBrown,SagawaGS96,NavratilBO97,TownerH02}, it
may influence precise model calculations depending on the importance of the
charge dependence in pairing correlations.

\section{Isospin Invariance Breaking and Isospin Mixing }

The estimate for the model parameters (Table \ref{tab:fitStat}) can determine
the extent to which the isospin symmetry is broken while $T$ remains a good
quantum number. Breaking of the isospin invariance
$\left|D/\frac{E}{2\Omega }\right|$ (Table \ref{tab:fitStat}) is in general
negligible for light nuclei (\dlevel and \flevel levels) in agreement with the
experimental  data. For medium mass nuclei in the \fpg major shell the isospin
breaking is significantly greater. Furthermore, as expected from observations,
for the \dlevel level the interaction strengths of all $pn$, $pp$ and $nn$
pairing are almost equal ($T$ is a good quantum number),
$F\approx 0$ (Table \ref{tab:fitStat}), and they differ for the
\flevel and for the
\fpg  shells, with the $pn$ isovector strength being more attractive,
$F>0$. Indeed, the $F$ isospin  mixing interaction strength is extremely small
and hence a  charge independent nuclear interaction (where $F$ is
neglected) comprises a quite reasonable approximation. The latter yields
major simplifications to the pairing problem and consequently most of isovector
pairing studies have been done assuming good isospin.

The question regarding how strong individual isospin non-conserving
nuclear interactions are
[such as (\ref{HINC})] still remains open -- there are no sharp
answers at the present level
of experimental results and microscopic theoretical interpretations.
It  is only their overall
contribution that is revealed by the free nucleon-nucleon data
\cite{Henley66} to be slightly (by $2\%$) more attractive in the $pn$
$T =1$ system than  the
$pp$ one. Within the framework of the \Spn{4} model, the charge
dependence of the pure
nuclear interaction can be estimated through the comparison of the
$T_0=0$ two-body model
interaction [(\ref{clH}) with $\varepsilon =0$] relative to the
$T_0=1$ in the $T  =1$
multiplets, which, for example in the \flevel level, is on average
$\sim 2.5\%$. This
estimation does not aim to confirm the  charge-dependence, which is
very difficult at this
level of accuracy compared to the  broad energy range considered in
the model for nuclei with
masses $32 \le A \le 100$. Nonetheless,  it reflects the fingerprints
of the experimental data
in the properties of the model  interaction (\ref{clH}).

\begin{figure}[th]
\centerline{\epsfxsize=3.0in\epsfbox{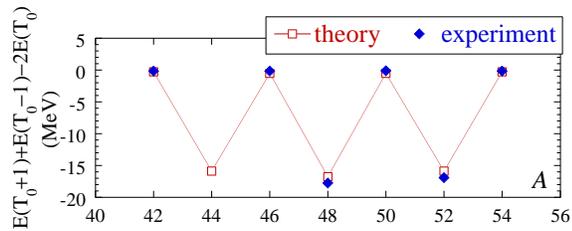}}
\caption{ Energy difference, $E(T_0+1)+E(T_0-1)-2E(T_0)$
with $T_0=0$, for the lowest \IASs~ in nuclei in the
\flevel level (or twice the $c$-coefficient of (\ref{IMME}) for the
$A/2$-odd $T=1$ multiplets)
according to the \Spn{4} model (red solid line with open squares) in comparison
to the experiment (blue diamonds). }
\label{Stg2ic}
\end{figure}

In addition, the \Spn{4} model reproduces reasonably well the
$c$-coefficient in the well-known isobaric multiplet mass equation
\cite{Wigner57,BenensonK79,Ormand97}
\begin{equation} a+bT_0+cT_0^2,
\label{IMME}
\end{equation}
for the binding energies of isobaric analogs (of the  same mass number
$A$, isospin $T$, angular momentum $J$, etc.), where the coefficient
$c$ ($b$) depends on the isotensor (isovector) component of the nuclear
interaction [i.e., of  rank 2 (1) with  respect to isospin `rotations'].
The $c$-coefficient is indeed an energy filter,
$2c=E(T_0+1)+E(T_0-1)-2E(T_0)$, for a given mass number $A$ and isospin $T$. In
the framework of our model, this energy function for the lowest 
\IASs~ was found
to be in a good agreement with  observed fine-structure effects (where data was
available) \cite{SGD03stg} and it reproduces the experimental 
staggering behavior
with respect to $A$ (Figure \ref{Stg2ic}). Both theoretical and experimental
results show that this finite energy difference when centered at an
$N=Z$ odd-odd nucleus ($T_0=0$ and $A/2$ odd), and hence $c$, is negative and
very close to zero for $T=1$ multiplets in the \flevel shell (see Figure
\ref{Stg2ic} for $A/2$ odd). Such an agreement of the \Spn{4} model 
outcome with
experimental evidence is a valuable result. The requirement that the
coefficients of (\ref{IMME}) are well reproduced is essential for the isospin
non-conserving models
\cite{TownerHH77,OrmandBrown,ZukerLMP02}, which has been achieved in
\cite{TownerHH77}  by increasing (approximately by $2 \%$) of all the
$T=1$ $pn$ matrix
elements relative to the $nn$ ones and which has lead to a conclusion
in \cite{ZukerLMP02}
that the  isotensor nature of the nuclear interaction is dominated by
a $J=0$ pairing term.

In short, the freedom allowed by introducing additional parameters
(as $F$ and $D$) reflects
the symmetries observed in light nuclei (good  isospin) and the
comparatively larger
symmetry-breaking as expected in medium-mass  nuclei. Hence, the
charge dependence of the
nuclear force, being a very challenging problem, yields results,
based on a simple
group-theoretical approach, that  are qualitatively consistent with
the observations.

\subsection{Near Isospin Symmetry of the Isobaric Analog $0^+$ States}

Both empirical evidence (such as scattering analysis and finite
energy differences) and the
comparison of the model to experimental data (Table
\ref{tab:fitStat}) does not yield equal pairing strengths ($F
\gtrapprox 0$) resulting in a coupling of isospin eigenstates $\left|
n,T,T_0\right\rangle $
from different isospin multiplets with a degree of mixing expected to be
very small. Therefore, the eigenvectors of the total Hamiltonian (\ref{clH}),
$\left| n(\tilde{T})T_0 \right\rangle$, have an almost good isospin
$\tilde{T}$ quantum
number. Their overlap with the states of definite isospin  values
yields an estimate for  the
magnitude of the isospin admixture (see Table
\ref{tab:isoMix} for the \dlevel and \flevel orbits):
\begin{equation}
\delta _{\tilde{T },T }=\left| \left\langle n,T ,T_0| n(\tilde{T
})T_0 \right\rangle \right| ^{2}*100[\%].
\label{delta}
\end{equation}
\begin{table}[th]
\caption{\Spn{4} model estimate for the overlap [\%] of \IASs~ of almost good
isospin $\tilde{T}$ with the states of definite isospin for $^{36}$Ar in the
\dlevel and the nuclei in the \flevel level. The table is symmetric 
with respect
to the sign of $n-2\Omega $.}
\center{
\begin{tabular}{cclllll}
\hline \hline
$^{A}$X$^{(\tilde T)}$  & $(N_{+1},N_{-1})$ & $T =0$ & $T =1$ & $T =2$ & $T =3$
&
$T =4$ \\
\hline
$^{36}$Ar$^{(0)}$ & $(2,2)$ & 99.9999 & - & 0.0001 & - & - \\
$^{44}$Ti$^{(0)}$ & $(2,2)$ & 99.90 & - & 0.10 & - & - \\
$^{46}$Ti$^{(1)}$ & $(2,4)$ & - & 99.98 & - & 0.02 & - \\
$^{46}$V$^{(1)}$ & $(3,3)$ & - & 99.98 & - & 0.02 & - \\
$^{46}$Cr$^{(1)}$& $(4,2)$ & - & 99.98 & - & 0.02 & - \\
$^{48}$Ti$^{(2)}$ & $(2,6)$ & - & - & 99.997 & - & 0.003 \\
$^{48}$V$^{(2)}$ & $(3,5)$ &  - & - & 99.994 & - & 0.006 \\
$^{48}$Cr$^{(0)}$ & $(4,4)$ &  99.83534 & - & 0.16465 & - & $10^{-5}$ \\
$^{48}$Cr$^{(2)}$ & $(4,4)$ &  0.143 & - & 99.849 & - & 0.008 \\
$^{48}$Mn$^{(2)}$ & $(5,3)$ &  - & - & 99.994 & - & 0.006 \\
$^{48}$Fe$^{(2)}$ & $(6,2)$ & - & - & 99.997 & - & 0.003 \\
\hline \hline
\end{tabular}
}
\label{tab:isoMix}
\end{table}

The overlap percentages in Table \ref{tab:isoMix} confirm that the nuclear
lowest \IASs~ have primarily isospin $T =\left| T_0\right| $ for even-even, and
$T =\left|T_0 \right| +1$ for odd-odd nuclei, with a very small
mixture of the higher possible
isospin values. As it is expected, the $\delta  _{\tilde{T },T } $
isospin mixing increases as
$Z$ and $N$ approach one another and towards the  middle of the
shell. For nuclei occupying a
single-$j$ shell, the mixing of the isospin states  is less than
$0.17\%$.  Although the isospin mixing is negligible for light nuclei
in the $ j=3/2$ orbit,
it is clearly bigger for the $ j=7/2$ level. The mixing  is expected
to be even stronger in
multi-shell configurations.

\subsection{Non-Analog $\beta $-Decay Transitions}

For a superallowed Fermi $\beta$-decay transition ($0^+ \rightarrow
0^+$) the $ft$ comparative
lifetime is nucleus-independent according to the
conserved-vector-current (CVC) hypothesis and
given by
\begin{equation}
ft=\frac{K}{G_V^2|M_F|^2 }, \quad K =2\pi ^3\hbar \ln 2
\frac{(\hbar c)^6}{(m_e c^2)^5},
\end{equation}
where $K/(\hbar c)^6=8.120 270 (12)\times 10^{-7}$ GeV$^{-4}$s ($m_e$
is the mass of the
electron) and $G_V$ is the vector coupling  constant for nuclear
$\beta $ decay (see for
example \cite{OrmandBrown}). $M_F$ is  the Fermi matrix element $\left<
\text{F}|\sqrt{2\Omega} T _{\pm } |\text{ I} \right>$  between a
final (F) state with isospin
projection $T _0^{\text{F}}$ and an initial (I)  states with $T
_0^{\text{I}}$ in a decay
generated by the raising (for
$\beta ^-$ decay) and lowering ($\beta ^+$) isospin transition
operator\footnote{The factor
of $\sqrt{2\Omega}$ appears  due to the normalization of the basis
operators adopted in the \spn{4}
algebraic model.} $\sqrt{2\Omega}T _{\pm }$, which in the framework
of our model is given as
\begin{equation}
|M_F|^2=2\Omega
|\left< \text{F};n(\tilde T)T_0\pm 1|T _{\pm }
|\text{I};n(\tilde T)T_0 \right>|^2.
\end{equation}
Typically, the isospin impurity caused by isospin non-conserving
forces in nuclei is estimated
as a correction to the Fermi matrix element
$|M_F|^2$ of the superallowed $\tilde T$ analog $0^+
\rightarrow 0^+$ transition,  $\delta _C= 1- |M_F|^2/\left\{\tilde T
(\tilde T +1) -T
_0^{\text{F}}T _0^{\text{I}} \right\}$. For more than two-state
mixing, the degree of isospin
admixture between \IASs~ should be estimated using  the normalized
transition matrix element
between non-analog (NA) states (e.g. \cite{TownerH02}),
\begin{equation}
\delta_{IAS}=\frac{|M_F^{\text{NA}}|^2}{\left\{ \tilde T (\tilde T
+1) -T _0^{\text{F}}T
_0^{\text{I}}\right\}},
\label{deltaIAS}
\end{equation}
where $\tilde T$ is the almost good isospin of the parent nucleus (see
Table \ref{tab:beta} for \flevel ). In general, the
$\delta _{IAS}$ correction may be very different than the
order of the $\delta _{\tilde{T },T }$ overlap quantity (\ref{delta})
presented in Table
\ref{tab:isoMix} because in decays the degrees of isospin mixing
between non-analog states
within both the parent and daughter nuclei are significant.

\begin{table}[th]
\caption{Non-analog $\beta $-decay transitions to energetically
accessible $0^+$ states under consideration and the
corresponding isospin mixing estimates $\delta _{IAS}$ (\ref{deltaIAS}),
in $\%$, according to the \Spn{4} model for nuclei in  the \flevel level.
}
\center{
\begin{tabular}{llll}
\hline \hline
\multicolumn{3}{c}{$\beta$-decay} & $\delta _{IAS},$ \\ [1ex]
$^A_ZX^{(\tilde T _X)}$ & $\rightarrow $ & $ ^{\hspace{0.35cm}
A}_{Z-1}Y^{(\tilde T _Y)}$ & $\%$ \\ [1ex]
\hline
$_{23}^{44}\text{V}^{(2)} $ & $\rightarrow $ & $ _{22}^{44}\text{Ti}^{(0)}$ &
0.098 \\ [1ex]
\hline
$_{25}^{46}\text{Mn}^{(3)} $ & $\rightarrow $ & $ _{24}^{46}\text{Cr}^{(1)}$
     &0.0169 \\ [1ex]
$_{24}^{46}\text{Cr}^{(3)} $ & $\rightarrow $ & $ _{23}^{46}\text{V}^{(1)}$
     &0.0104 \\ [1ex]
$_{23}^{46}\text{V}^{(3)} $ & $\rightarrow $ & $ _{22}^{46}\text{Ti}^{(1)}$
     &0.00447 \\ [1ex]
\hline
$_{27}^{48}\text{Co}^{(4)} $ & $\rightarrow $ & $ _{26}^{48}\text{Fe}^{(2)}$
     &0.00327 \\ [1ex]
$_{26}^{48}\text{Fe}^{(4)} $ & $\rightarrow $ & $ _{25}^{48}\text{Mn}^{(2)}$
     &0.00280 \\ [1ex]
$_{25}^{48}\text{Mn}^{(4)} $ & $\rightarrow $ & $ _{24}^{48}\text{Cr}^{(2)}$
&0.00189 \\[1ex]
$_{24}^{48}\text{Cr}^{(4)} $ & $\rightarrow $ & $ _{23}^{48}\text{V}^{(2)}$
&0.00103 \\[1ex]
$_{23}^{48}\text{V}^{(4)} $ & $\rightarrow $ & $ _{22}^{48}\text{Ti}^{(2)}$
&0.00038 \\[1ex]
$_{25}^{48}\text{Mn}^{(4)} $ & $\rightarrow $ & $ _{24}^{48}\text{Cr}^{(0)}$
&$4.5 \times 10^{-7}$ \\[1ex]
$_{25}^{48}\text{Mn}^{(2)} $ & $\rightarrow $ & $ _{24}^{48}\text{Cr}^{(0)}$
&0.14328 \\[1ex]
\hline
$_{27}^{50}\text{Co}^{(3)} $ & $\rightarrow $ & $ _{26}^{50}\text{Fe}^{(1)}$
     & 0.0169 \\ [1ex]
$_{26}^{50}\text{Fe}^{(3)} $ & $\rightarrow $ & $ _{25}^{50}\text{Mn}^{(1)}$
     &0.0104 \\ [1ex]
$_{25}^{50}\text{Mn}^{(3)} $ & $\rightarrow $ & $ _{24}^{50}\text{Cr}^{(1)}$
     &0.00447 \\ [1ex]
\hline
$_{27}^{52}\text{Co}^{(2)} $ & $\rightarrow $ & $ _{26}^{52}\text{Fe}^{(0)}$
     &0.098 \\ [1ex]
\hline \hline
\end{tabular} }
\label{tab:beta}
\end{table}

The analysis of the results shows that the mixing between \IASs~
(which is at least
$\Delta T =2$ mixing) is on average $0.006 \%$ excluding even-even
$N=Z$ nuclei. This is on the order of a magnitude less than the
mixing of the first excited
$0^+$ non-analog  state due to isospin non-conserving interaction,
which is  typically about
$0.04 \%$ for the
\flevel level \cite{Hagberg,TownerH02}. In addition, this yields nonanalog
$\beta$-decays weaker than possible Gamow-Teller transitions; the strength of
the latter is found to be less than $0.02\%$ of the total $\beta$-decay strenght
for the nuclei in the
\flevel shell \cite{Hagberg} and to substantially increase with increasing mass
number $A$ \cite{Hamamoto93,HardyT02,Piechaczek03}. This makes
$\delta _{IAS}$ mixing very difficult to be detected especially when
the isospin-symmetry
breaking correction ($\delta _C$) to analog Fermi  matrix elements in
this level is on the
order of a percent \cite{DobaczewskiH95,HardyT02}.

\begin{figure}[th]
\centerline{\epsfxsize=3.4in\epsfbox{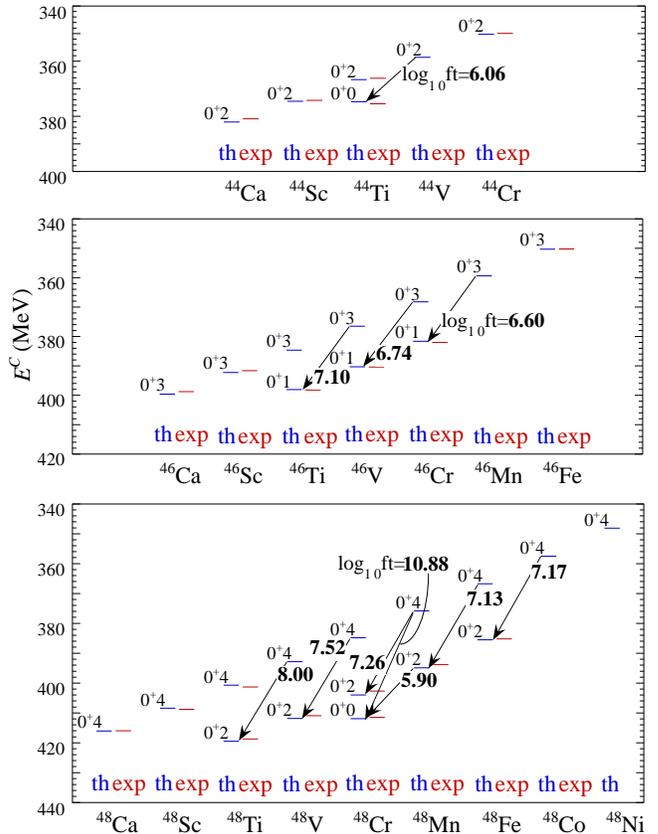}}
\caption{  Non-analog $0^+ \rightarrow 0^+$ $\beta  ^+$-decay transitions to
energetically accessible $J^\pi \tilde T$ states under consideration
indicated by arrows and by the corresponding theoretically calculated $\log
_{10}ft$ values (there are no available experimental $ft$ values for
comparison). The theoretical (blue, ``th") and experimental
\cite{AudiWapstra,Firestone} (red, ``exp") binding energies $E^C$ in MeV
(including the Coulomb potential energy) are shown for the isobar 
sequences with
$A=44$ to $A=48$ in the \flevel level.}
\label{CaBetaDecays}
\end{figure}

Not surprising, the largest values for the $\delta _{IAS} $
correction are observed for $\Delta T=2$
$\beta ^{\pm}$-decays to energetically accessible $0^+$ ground states of
even-even
$N=Z$ nuclei (Table \ref{tab:beta}). While for these decays $\delta
_{IAS}$ is extremely
small, namely less than $0.14 \%$, as expected for the  contribution
of the higher-lying $0^+$
states \cite{TownerH02}, it is comparable to the  order of
isospin-symmetry breaking
corrections for the \flevel orbit that are typically taken  into account
\cite{TownerH02}. The reason may be that for the even-even $N=Z$
nuclei the second-lying \IASs~ are situated relatively low due to a significant
$pn$  interaction (Figure \ref{CaBetaDecays}).

Above all, the $\delta _{IAS} $ results in Table \ref{tab:beta} 
clearly show the
overall pattern and the order of significance of the isospin mixing under
consideration. This is evident within the first-order approximation in terms of
the $F$ parameter ($F \ll 1$) of $\delta _{IAS} $, which for
\flevel deviates on average by only 2\% from its exact  calculations in Table
\ref{tab:beta}. The $\delta _{IAS} $ isospin mixing correction is then
proportional to $F^2$ and one finds out that its order of  magnitude 
remains the
same for large variations of the $F$ parameter of more than 60\%. In addition,
greater $F$ values are not very likely because the $\delta
_{IAS} $ estimates (Table \ref{tab:beta}) fall close below an upper
limit, which does not contradict experimental and theoretical results for other
types of isospin mixing.

Moreover, in this first-order approximation the
ratio of any two isospin corrections for \flevel is independent of the
parameters of the model interaction. This implies that such a ratio does not
reflect at all the uncertainties of the interaction strength parameters but
rather it is  characteristic of the relative strength of both decays. It
identifies the  decay, for which the maximum isospin mixing correction is
expected in the \flevel orbit, namely
$_{25}^{48}\text{Mn}^{(2)} \rightarrow  _{24}^{48}\text{Cr}^{(0)}$, and as well
as the amount by which $\delta _{IAS} $ of the other possible non-analog decays
is relatively suppressed. For example, the $\delta _{IAS} $ correction for the
$_{23}^{44}\text{V}^{(2)}
\rightarrow  _{22}^{44}\text{Ti}^{(0)}$ decay is around two-thirds
the maximum one and that for the
$_{25}^{46}\text{Mn}^{(3)} \rightarrow  _{24}^{46}\text{Cr}^{(1)}$ decay is
around one-eighth the maximum one (Table
\ref{tab:beta}). Such a ratio quantity exhibits a general trend of increasing
$\delta _{IAS} $ isospin mixing with $Z$ within same isospin multiplets and as
well it reveals enhanced $\Delta T=2$ decays to the ground state of   even-even
$N=Z$ nuclei with increasing $\delta _{IAS} $ towards the middle of the shell.
Furthermore, the ratio retains its behavior for the non-analog
$\beta $ decays between nuclei with the same valence proton and neutron numbers
as in Table
\ref{tab:beta} but occupying the
\fpg major shell. Therefore, among the non-analog $\beta $ decays for the
$A=60-64$ isobars with valence protons and neutrons in the \fpg shell the
$\delta _{IAS} $ isospin mixing of the
$_{33}^{64}\text{As}^{(2)} \rightarrow  _{32}^{64}\text{Ge}^{(0)}$ decay is
expected to be the largest with a tendency of a further increase towards the
middle of the shell. In short, the significance of the isospin mixing caused by
a charge dependent
$J=0$ pairing correlations is evident from Table \ref{tab:beta} for the \flevel
level and continues this trend for the upper $fp$ shell.

The small mixing of the $0^+$ isospin eigenstates from different isospin
multiplets yields very small but nonzero $|M_F^{NA}|^2$  matrix elements for
non-analog $\beta ^{\pm} $ decay transitions as indicated by
$\delta _{IAS}$ in Table \ref{tab:beta}. For nuclei in the \flevel shell, such
non-analog transitions to energetically accessible states are shown in Figure
\ref{CaBetaDecays} along with the $ft$  values (where we use $K/G_V^2=6200s$
\cite{Gerhart54and58}).  In the framework of the \Spn{4} model, these 
values are
symmetric with respect to the sign of $T_0$ [possible $\beta ^-$ decays are not
shown in Figure
\ref{CaBetaDecays}] and of $n-2\Omega $ (the $A=50$ and $A=52$ multiplets are
analogous to the $A=46$ and $A=44$ ones, respectively). The results yield that
ten of the transitions are classified as forbidden ($\log _{10}ft \ge 
7$), other
four are suppressed ($\log _{10}ft \approx 7$) and the four decays to 
the ground
state of an even-even
$N=Z$ nucleus appear to have  comparatively larger decay rates ($\log _{10}ft
\approx 6$). In Figure \ref{CaBetaDecays} the theoretically calculated \IAS~
energies are shown together with the available experimental ones. It is worth
mentioning that while the energies of the lowest \IASs~ determined directly the
parameters of the model interaction, a quite good reproduction of the
experimental higher-lying
\IAS~energies followed without any parameter adjustment \cite{SGD04IAS}. This
outcome is important because the energy difference between two \IASs~ within a
nucleus directly affects the degree of their mixing. In  summary, the
theoretical \Spn{4}  model suggests the possible existence, albeit  highly
hindered, of
$\Delta T =2$ non-analog $\beta
$-decay transitions.

\section{Conclusions}

We employed a group-theoretical approach based on the \Spn{4} 
dynamical symmetry
to describe microscopically a possible isospin mixing induced by a short-range
charge dependent nuclear interaction. The \Spn{4} model interaction 
incorporates
the main driving forces, including $J=0$ pairing correlations and implicit
quadrupole-quadrupole term, that shape the nuclear pairing-governed
\IASs~in the \flevel  level where the \Spn{4} Hamiltonian correlates strongly
with realistic interactions. This approach provides a reasonable reproduction
not only of the energies of the lowest \IASs~in total of 149 nuclei. It also
reproduces the available experimental energies of the higher-lying \IASs~in the
\flevel  level and fine structure effects without any variation of the
parameters of the  model interaction. In this respect, as predicted by our
model, the coefficient related to the isotensor part  of a general
non-conserving force, $c$, which has been recently found to be dominated by a
charge dependent $J=0$ pairing interaction
\cite{ZukerLMP02},  agrees quite well with the experimental values.

The isospin-symmetry breaking due to coupling  of \IASs~ in nuclei was
estimated to be extremely
small for nuclei in the \dlevel and \flevel  orbitals with the $N=Z$
even-even nuclei being
an exception. For these nuclei, strong pairing correlations,
including a significant $pn$
interaction, are  responsible for the existence of comparatively
larger isospin mixing,
although the latter is still at least an order of a magnitude
smaller than
the overall isospin admixture in the ground state. The results
also show that
a variation of more than 60\% in the $F$ isospin mixing parameter is
required to reduce the present $\delta _{IAS}$ results by an order of a
magnitude.

The analysis also shows that there is a trend of
increasing isospin mixing  between \IASs~due to a charge dependent 
$J=0$ pairing
interaction towards the middle of the shell and for $\Delta T=2$ decays to the
ground state of an even-even $N=Z$ daughter nucleus. Such behavior is 
free of the
uncertainties in the strength parameters of the interaction
and is adequate for larger multi-$j$ shell domains such as
$\fpg$. For nuclei with valence protons and neutrons occupying
the \flevel
level the strongest non-analog decay is identified to be
$_{25}^{48}\text{Mn}^{(2)}
\rightarrow  _{24}^{48}\text{Cr}^{(0)}$, while the $\delta _{IAS}$ isospin mixing
correction for the rest of the decays is 2/3 to 1/300 the maximum one.

In short, the \spn{4}  algebraic model
yields an estimate for the  decay rates of possible
non-analog $\beta$-decay transitions due to a pure strong
interaction, which, though few of them may
affect slightly precise calculations, are not
expected to comprise the dominant contribution to the  isospin-symmetry
breaking correction tested in studies of superallowed Fermi $\beta$-decay
transitions.

\section*{Acknowledgments}
This work was supported by the US National Science  Foundation, Grant
Number 0140300. The
authors thank Dr. Vesselin G. Gueorguiev for his computational
{\small MATHEMATICA}\ programs
for non-commutative algebras.

\end{document}